\documentclass[fleqn,twoside,twocolumn,nofootinbib,showkeys]{revtex4} % Specifies the document class %,unsortedaddress
\usepackage[sec,nocpr]{ujp} % \usepackage[cyr]{ujp} for cyrillic, \usepackage[web]{ujp} for web
\begin{document}
\title[Stick-Slip Mode of Boundary Friction]%колонтитул
{STICK-SLIP MODE OF BOUNDARY FRICTION\\ AS THE FIRST-ORDER PHASE TRANSITION}%
\author{I.A.~Lyashenko}%1 автор
\affiliation{Sumy State University}%институт
\address{2, Rimskii-Korsakov Str., Sumy 40007, Ukraine}%адрес
\email{\!nabla04@ukr.net,\,zaskoka23@ukr.net}%e-mail
\author{A.M.~Zaskoka}%
\affiliation{Sumy State University}%
\address{2, Rimskii-Korsakov Str., Sumy 40007, Ukraine}%
%\email{nabla04@ukr.net, zaskoka_anton@mail.ru}%
\udk{533.9} \pacs{05.70.Ce; 05.70.Ln;\\[-3pt] 47.15.gm; 62.20.Qp;\\[-3pt]
64.60.-i; 68.35.Af; 68.60.-p} \razd{\secix}

\keywords{ultrathin lubricant film, boundary mode of friction,
tribological system} \autorcol{I.A.\hspace*{0.7mm}Lyashenko,
A.M.\hspace*{0.7mm}Zaskoka}

\setcounter{page}{91}%

\begin{abstract}
A tribological system consisting of two contacting blocks has been
considered. One of them is arranged between two springs, the other
is driven periodically. The kinetics of the system has been studied
in the boundary friction mode, when an ultrathin lubricant film is
contained between the atomically smooth surfaces. In order to
describe the film state, the expression for the free energy density
is used in the form of an expansion in a power series in the order
parameter, the latter being reduced to the shear modulus of
a lubricant. The stick-slip mode is shown to be realized in a wide
range of parameters, being a result of the periodic first-order phase
transitions between kinetic friction regimes. The behavior of the
system governed by internal and external parameters has been
predicted.
\end{abstract}

\maketitle

\section{Introduction}

Owing to the rapid development of high-precision experimental
techniques aimed at researching the nanosystems, the processes of
friction in the case where the thickness of a lubricant between
rubbing surfaces equals several atomic diameters have been
intensively studied recently~\cite{Persson,PopovBook,Isr-rev}. In
tribology, this friction mode was called \textquotedblleft boundary
friction\textquotedblright. It is often realized in ordinary
mechanisms, because the rubbing surfaces contact with each other by
means of surface irregularities or
inhomogeneities~\cite{Persson,AmJPhys}. The boundary friction
differs essentially from the hydrodynamic mode, when the friction
force is proportional to a power function of the velocity. Note that
the ultrathin film of a lubricant does not form conventional,
thermodynamically equilibrium phases, solid and liquid ones.
Instead, we have liquid- and solid-like states, which are kinetic
friction modes, and there can be several of
them~\cite{Yosh,Yosh-1996}. This occurs because the symmetry of a
lubricant state is substantially affected by friction surfaces, this
fact being not of importance for bulk lubricants. In the course of
friction, the phase transitions of both the first and second orders
can take place between the stationary states~\cite{Popov,1-order}.
These transitions often comprise the origin of the stick-slip motion
mode for contacting surfaces~\cite{Yosh,Filippov,Yosh-1996}.

In order to describe the boundary mode of friction and nano-contact
phenomena, phenomenological models are widely
used~\cite{Popov,Filippov,1-order,Carlson}. In particular, a model
was developed \cite{JPS}, in which the lubricant melting is driven by the thermodynamic
and shear mechanisms. In the framework of this model, the influence of
additive fluctuations of principal quantities was
studied~\cite{dissipative}, and their presence in the system was demonstrated to result in the
emergence of new stationary states and new kinetic friction
modes~\cite{UFJ,3-new}, which are not essential for bulk systems. The origin of
the hysteretic behavior, which was observed
experimentally~\cite{Isr-rev,exp1,exp-n}, was elucidated in work~\cite{FTT}. The indicated model
also made it possible to describe the periodic stick-slip mode of
motion~\cite{period,FrictWear-2010}.

In works~\cite{Popov,2000-SolStCom-Popov}, a thermodynamic scenario
of boundary friction was proposed, which is based on the phase
transition theory developed by Landau~\cite{stat}. This model takes
into account that the ultrathin film of a lubricant can melt and
stay in a liquid-like disordered state both owing to the ordinary
thermodynamic melting and as a result of overcoming the fluidity
threshold by the shear stress component (\textquotedblleft shear
melting\textquotedblright ). The influence of those factors was also
studied in work~\cite{1-order}, in which the excess
volume~\cite{Anael1,Anael2} arising owing to the lubricant
stochastization at its melting was selected as the order parameter.
As the excess volume increases, the shear modulus
decreases~\cite{1-order}, which results in the melting. In
works~\cite{Popov,2000-SolStCom-Popov}, the shear modulus itself was
selected as the order parameter, which acquires zero values in the
liquid-like phase. However, in
works~\cite{Popov,2000-SolStCom-Popov}, the melting is described as
a continuous phase transition of the second order, whereas jump-like
phase transitions of the first order are often observed in the
boundary friction mode~\cite{Yosh,Yosh-1996,1-order}, which are
responsible for the stick-slip motion~\cite{Yosh,Yosh-1996}.

This work is aimed at describing the first-order phase transition in the
framework of the model developed in works~\cite{Popov,2000-SolStCom-Popov}
and at studying the behavior of tribological systems on the basis of
the indicated modification. The model proposed does not make allowance for
specific types of lubricants, because its task consists in describing the
origins of phenomena that take place at the boundary friction. For specific
types of lubricants and friction surfaces, the model should be modified. To
some extent, it can be made by choosing the numerical values of coefficients
in the series expansions of the free energy, relaxation times, and so on. The
model describes only homogeneous lubricants composed of non-polar
quasispherical molecules~\cite{Yosh,Yosh-1996}. One of the reasons is the
fact that we study a situation where the elastic stresses acquire zero values in
the liquid-like state, i.e. the melting gives rise to the total disordering of
lubricant molecules, which does not takes place in thin lubricant films
consisting of polymer molecules. Another reason is the fact that the
obtained time dependences of the friction force and stresses are strictly
periodic, which is also observed for quasispherical molecules
only~\cite{Yosh,Yosh-1996}.

%Fig. 1
\begin{figure}% figure* for wide figure, [h] [!] to change the placement
\includegraphics[width=\column]{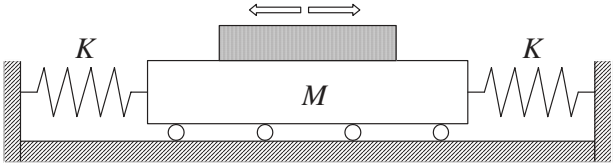}
\vskip-3mm\caption{Diagram of a tribological system  }
\end{figure}

\section{Tribological System}

The fabrication of atomically smooth surfaces with large dimensions is
associated with considerable technological difficulties. Therefore, to
measure the dynamic parameters of ultrathin lubricant films between such
surfaces, the surfaces characterized by small dimensions and pasted on spherical
or cylindrical surfaces that rub each other are used. This scheme was
applied while designing the surface force apparatus
(SFA)~\cite{Isr-rev,Isr1,Isr2}. Two types of SFA---Mk~II and Mk~III---were described in
review~\cite{Isr-rev}. In the latter, the system to control distances
between rubbing surfaces was improved. The device allows the shape of
surfaces to be determined, as well as the distance between them to within an
accuracy of 1~\AA . The contact area between surfaces is measured with an
accuracy of $\pm 5$\%, the normal and shear components of operating forces
to within $\pm 1$\%, and the magnitude of applied loading to within $\pm
5$\%.

One of the rubbing surfaces in the SFA is fixed, and the other is
driven to move periodically. In the course of motion, the shear
stresses and the effective viscosity of a lubricant are measured,
the lubricant structure is determined, and so forth. In this work,
we consider a simplified mechanical analog of the SFA exhibited in
Fig.~1. Two springs characterized by the stiffness constant $K$ are
connected with a block of mass $M$ mounted on rollers. The rolling
friction for the latter is neglected below. On the indicated block,
another block is arranged, which is brought into a periodic motion
by applying an external force. Provided that the surfaces of two
blocks interact with each other, the motion of the upper block
stimulates the motion of the lower one. The trajectory of the lower
block substantially depends on the friction mode established in the
system. A similar tribological system was experimentally studied in
works~\cite{TribInt-1999,TribInt-1997}. Note that, in contrast to
the SFA design, now both blocks are mobile, which enables the time
dependences of block coordinates and velocities to be registered,
and, by analyzing them, the rheological and tribological
characteristics of the system to be determined.\looseness=1

Let $X$ and $V=\dot{X}$ be the coordinate and the velocity, respectively, of
the upper block, whereas $x$ and $v=\dot{x}$\ denote the corresponding
quantities for the lower block. Let us consider the case where the upper
block moves according to the cyclic law,
%1
\begin{equation}
X=X_m \cos\omega t, \label{eq_1}
\end{equation}\vspace*{-5mm}
%2
\begin{equation}
 V=-X_m\omega\sin\omega t,
\label{eq_2}
\end{equation}
where $X_{m}$ is the amplitude, and $\omega $ is the cyclic frequency.
We write down the equation of motion for the lower block in the
form~\cite{TribInt-1999}
%3
\begin{equation} M\ddot{x}+2Kx-F=0,
\label{eq-3}
\end{equation}%
where $F$ is the friction force that arises between the blocks at their
relative motion. From the last expression, it follows that the character of
motion in the system essentially depends on the friction mode and the
lubricant properties, because they determine the force $F$.

The friction force is determined in a standard way,
%4
\begin{equation}
F=\sigma A,   \label{eq-4}
\end{equation}
where $\sigma$ is the shear stress that arises in the lubricant, and $A$ is
the contact area between rubbing surfaces.

In the boundary friction mode, the elastic, $\sigma _{\rm el}$, and
viscous (dissipative), $\sigma _{v}$, stresses arise in the
lubricant layer~\cite{Popov,1-order,exp-n}. As a rule, the melting is
accompanied by a reduction of the elastic stress component, whereas
the viscous one grows owing to an increase of the relative shear
velocity between rubbing surfaces~\cite{exp-n}. Hence, the total
stress is determined by the sum of indicated components,
%5
\begin{equation}
\sigma =\sigma _{\rm el}+\sigma _{v}.  \label{eq-5}
\end{equation}%
The viscous stresses in the lubricant layer are determined by the empirical
formula~\cite{Wear,Perss-PRL}
%6
\begin{equation}
\sigma _{v}=\frac{\eta _{\rm eff}(V-v)}{h},  \label{eq-6}
\end{equation}%
where the effective viscosity of the lubricant, $\eta _{\rm eff}$ (it
depends on plenty of factors and is determined experimentally) is
introduced into consideration, as well as the relative velocity of
surface motion, $V-v$.

As a result, in the case of boundary friction, polymer solutions or melts
are applied as lubricants. The necessity of such an application is caused by
the fact that the friction surfaces are small in dimensions, and the lubricant
film between them must not be squeezed out under the influence of large
tribological loadings. Such lubricants are non-Newtonian fluids, the
viscosity of which depends not only on the temperature, but also on the
velocity gradient. However, the application of the SFA allows the behavior of a
wide class of lubricants to be examined in the boundary friction mode,
because the rubbing surfaces in these experiments are completely imbedded
into a vessel with a liquid to study, so that the latter is not squeezed out
from the gap between the surfaces during their motion~\cite{Isr-rev}.
However, we should note that even ordinary water, when being used as a
boundary lubricant, can behave as a non-Newtonian fluid, because, owing to
its interaction with the surfaces, it can create spatially ordered
structures in the course of motion.

The non-Newtonian fluids are divided into two classes: pseudoplastic fluids,
the viscosity of which decreases with the growth of the strain rate (e.g.,
these are polymer solutions and melts) and dilatant ones, the viscosity of
which increases as $\dot{\varepsilon}$ grows (e.g., suspensions of solid
particles). For both situations to be taken into account, let us use a
simple power-law approximation~\cite{Wear,Perss-PRL}
%7
\begin{equation}
\eta _{\rm eff}=k(\dot{\varepsilon})^{\gamma }.  \label{eq-7}
\end{equation}%
Here, we introduced the proportionality coefficient $k$ (its
dimension is $\mathrm{Pa}\cdot \mathrm{s}^{\gamma +1}$) and the
dimensionless index $\gamma $ (for pseudoplastic fluids, $\gamma
<0$; dilatant ones are characterized by the index $\gamma >0$; and
$\gamma =0$ in the case of Newtonian fluids).

The strain rate is determined through the relative velocity of motion and
the lubricant thickness $h$~\cite{Wear},
%8
\begin{equation}
\dot{\varepsilon}=\frac{V-v}{h}.  \label{eq-8}
\end{equation}%
Taking Eqs.~(\ref{eq-7}) and (\ref{eq-8}) into account, the expression for
viscous stresses (Eq.~(\ref{eq-6})) looks like
%9
\begin{equation}
\sigma _{v}=k\left(\! \frac{V-v}{h}\!\right) ^{\gamma +1}\!.
\label{eq-9}
\end{equation}%
Note that, according to Eq.~(\ref{eq-9}), viscous stresses are available in
both the liquid- and solid-like states of a lubricant. The presence of
viscous (dissipative) stresses in both phases was indicated in the experimental
work~\cite{exp-n}. However, if the lubricant is in a solid-like state,
viscous stresses are low, because, in accordance with Eq.~(\ref{eq-9}), they
are proportional to the relative shear velocity, $V-v$, which is low in this
case.

Substituting Eqs.~(\ref{eq-5}) and (\ref{eq-9}) into Eq.~(\ref{eq-4}), we
obtain the final expression for the friction
force~\cite{JtfL2,JtfL3,LKM-upj-2011,TribInt},
%10
\begin{equation}
F=\left[ \sigma _{\rm el}+k\ \mathrm{sgn}(V-v)\left(\!
\frac{|V-v|}{h}\!\right) ^{\gamma +1}\right] A,  \label{F}
\end{equation}where the function
%11
\begin{equation}
\mathrm{sgn}(V-v)=\left\{\!\! \begin{array}{c@{\;}c}
1, & V\geq v, \\[0.2cm]
-1, & V<v\end{array}\right.   \label{eq-11}
\end{equation}%
takes into account the direction of force action. The first term
in Eq.~(\ref{F}) describes the elastic component of the friction force,
and the second is responsible for the viscous one, which grows with
the velocity. Hence, the friction force depends on the
velocity of the lower block, $v$, and the elastic stresses, $\sigma
_{\rm el}$, that arise in a lubricant.

\section{Thermodynamic Model}

%Fig. 2
\begin{figure}[b]% figure* for wide figure, [h] [!] to change the placement
\includegraphics[width=7.0cm]{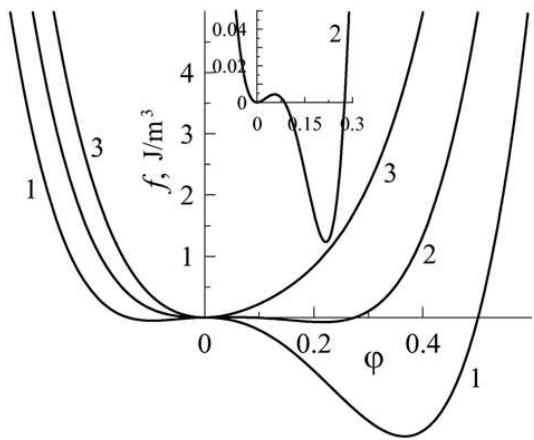}
\vskip-3mm\caption{Dependences of the free energy density $f$ (see
Eq.~(\ref{eq-12})) on the dimensionless order parameter $\varphi$
for various temperatures $T=265$, 286, and 310$~\mathrm{K}$ (curves
\textit{1} to \textit{3}, respectively). The calculation parameters
are $\alpha=0.95~\mathrm{J/(K \cdot m^{3})}$, $T_{c}=290$~K,
${a=4\times10^{12}}$~Pa, $b=230$~J/m$^{3}$, $c=850$~J/m$^{3}$, and
the shear strain $\varepsilon _{\rm el}=2.1\times10^{-6}$
}\vspace*{1.5mm}
\end{figure}

In the homogeneous case, the free energy density for an ultrathin lubricant layer
looks like~\cite{Popov,JtfL2,JtfL3,2000-SolStCom-Popov}\
%12
\begin{equation}
f=\alpha (T-T_{c})\varphi ^{2}+\frac{a}{2}\varphi ^{2}\varepsilon
_{\rm el}^{2}-\frac{b}{3}\varphi ^{3}+\frac{c}{4}\varphi ^{4},
\label{eq-12}
\end{equation}%
where $T$ is the lubricant temperature; $T_{c}$ is the critical
temperature; $\varepsilon _{\rm el}$ is the shear component of the
elastic strain; $\alpha $, $a$, $b$, and $c$ are positive constants,
and $\varphi $ is the order parameter (the amplitude of the periodic
component in the microscopic medium density
function~\cite{Popov,2000-SolStCom-Popov}). The parameter $\varphi $
equals zero in the liquid-like phase and acquires nonzero values in
the solid-like one. In comparison with
works~\cite{Popov,2000-SolStCom-Popov}, potential~(\ref{eq-12})
takes additionally the third-order term into account. This form of
expansion is used to describe phase transitions of the first
order~\cite{stat,Popov1}. In the second term in Eq.~(\ref{eq-12}),
we also introduced the factor $a$, which allows us to vary the
contribution of the elastic energy to the potential.

The elastic stresses that arise in the lubricant layer, according to
Eq.~(\ref{eq-12}), are determined as the derivative $\sigma _{\rm
el}=\partial f/\partial \varepsilon _{\rm el}$, so that
%13
\begin{equation}
\sigma _{\rm el}=a\varphi ^{2}\varepsilon _{\rm el}.  \label{eq-13}
\end{equation}%
Therefore, after the coefficient $a$ has been introduced into expansion
(\ref{eq-12}), the shear modulus is determined as follows:
%14
\begin{equation}
\mu =a\varphi ^{2}.  \label{eq-14}
\end{equation}%
Depending on the value of parameter $a$, it can acquire both small
and large values at $\left\vert \varphi \right\vert <1$. Note that,
in the boundary friction mode, the shear modulus can be several
orders of magnitude larger than that in the hydrodynamic mode for
the same lubricant. As a result, if the critical temperature $T$ or
critical elastic shear stress $\sigma _{\rm el}$ become exceeded in
the course of friction, the lubricant does not melt completely;
instead, a domain structure with regions of liquid-assisted and dry
friction is created. For this situation to be studied,
Eq.~(\ref{eq-12}) must include gradient terms, which considerably
complicates the subsequent consideration. However, the examination
of such spatial structures comprises a separate problem, which is
not the purpose of this work. Therefore, the gradient terms are
excluded from Eq.~(\ref{eq-12}), which corresponds to the
consideration of a lubricant behavior in the framework of the
one-domain model with a homogeneous \mbox{structure}.\looseness=1

According to the principle of minimum energy, the system tends to occupy a
stationary state, which corresponds to the minimum of the free energy $f(\varphi
)$ (see Eq.~(\ref{eq-12})), irrespective of its initial conditions. Since
the parameter $\varphi $ is the amplitude of the periodic component in the
microscopic medium density function, we consider below only the physical
range of values $\varphi \geq 0$. Let us introduce the
function
%15
\begin{equation}
B(\varepsilon _{\rm el},T)=a\varepsilon _{\rm el}^{2}+2\alpha
(T-T_{c}). \label{B-funk}
\end{equation}%
The analysis of expression~(\ref{eq-12}) for the free energy allows
the following situations to be distinguished. Provided that the
condition ${B(\varepsilon _{\rm el},T)\leq 0}$ is obeyed, the
maximum of potential~(\ref{eq-12}) at $\varphi =0$ and its minimum
at $\varphi >0$ are realized (curve~\textit{1} in Fig.~2). In this
case, the lubricant is solid-like, because the shear modulus $\mu
>0$. In the intermediate interval ${0<B(\varepsilon _{\rm
el},T)<{b^{2}}/(4c),}$ the maximum of the potential at $\varphi =0$
transforms into a minimum and, additionally, there emerges a maximum
that separates the zero and nonzero minima (curve~\textit{2} in
Fig.~2; it is also shown scaled-up in the inset). In this case, the
state of a lubricant depends on the initial conditions, and the lubricant
can be in either solid- or liquid-like state. In the latter case,
$B(\varepsilon _{\rm el},T)\geq b^{2}/(4c)$, and a single minimum of
the potential at $\varphi =0$ is realized (curve~\textit{3} in Fig.~2),
which, according to Eq.~(\ref{eq-14}), corresponds to the zero value
of lubricant shear modulus and its liquid-like structure.

The stationary values of order parameter $\varphi $ are determined
as the roots of the equation \mbox{$\partial f/\partial \varphi =0$}
\cite{JtfL2,JtfL3}, namely,
%16
\begin{equation}
\varphi _{\mp }=\frac{b}{2c}\mp \sqrt{\left( \!\frac{b}{2c}\!\right)
^{2}-\left(\! \frac{a}{c}\varepsilon _{\rm el}^{2}+\frac{2\alpha
(T-T_{c})}{c}\!\right) }. \label{eq-15}
\end{equation}%
The root $\varphi _{-}$ is related to the unstable stationary state,
because it corresponds to the maximum of potential~(\ref{eq-12}).
The stable state, which corresponds to the potential minimum, is
given by the root $\varphi _{+}$. Besides roots~(\ref{eq-15}), the
stationary solution $\varphi _{0}=0$ always exists, which
corresponds to the extremum of potential~(\ref{eq-12}) with the order
parameter equal to zero; it can be either a maximum or minimum of
the potential. According to Eq.~(\ref{eq-15}), the lubricant melts
if either the temperature $T$ elevates or the shear component of
the elastic strain, $\varepsilon _{\rm el}$, grows. Thus, the model
concerned makes allowance for both thermodynamic and shear meltings.

As was already indicated above, at small values of temperature $T$
and strain $\varepsilon _{\rm el}$, when the function $B(\varepsilon
_{\rm el},T)\leq 0$, the lubricant is solid-like, because, in
accordance with Eq.~(\ref{eq-15}), a stationary value of parameter
$\varphi $ different from zero is realized, and, according to
Eq.~(\ref{eq-14}), the shear modulus $\mu $ is also non-zero. In
this case, the potential has a single minimum at $\varphi \geq 0$.
If the temperature $T$ exceeds the critical value
%17
\begin{equation}
T_{c0}=T_{c}-\frac{a}{2\alpha }\varepsilon _{\rm
el}^{2}+\frac{b^{2}}{8\alpha c}, \label{eq-16}
\end{equation}%
the order parameter vanishes in a jump-like manner, when the lubricant
passes into the liquid-like state, in which the potential $f(\varphi )$
has a single minimum at $\varphi =0$~\cite{JtfL2,JtfL3}. If, after this
transition, the temperature $T$ falls down further, the lubricant solidifies
following the mechanism of first-order phase transformation at a lower
temperature,
%18
\begin{equation}
T_{c}^{0}=T_{c}-\frac{a}{2\alpha }\varepsilon _{\rm el}^{2},
\label{eq-17}
\end{equation}%
and the parameter $\varphi $ becomes non-zero again. In the
intermediate temperature region, ${T^{0}_c<T<T_{c0}}$, the potential
is characterized by two minima at positive $\varphi$. Hence, the
dependence $\varphi (T)$ has a hysteresis
character~\cite{JtfL2,JtfL3} and corresponds to the phase transition
of the first order. Expression~(\ref{eq-17}) elucidates the physical
meaning of the critical temperature $T_{c}$; namely, it is the
temperature of lubricant solidification at zero strains, when only
the mechanism of thermodynamic melting is active in the system.

From expression~(\ref{eq-16}), it follows that the lubricant melts not only
at the temperature elevation, but also if it is subjected to an external
mechanical action, when the elastic strain component exceeds the critical
value
%19
\begin{equation}
\varepsilon _{{\rm el},c0}=\sqrt{\frac{2\alpha
(T_{c}-T)}{a}+\frac{b^{2}}{4ac}}. \label{eq-18}
\end{equation}%
Using formula~(\ref{eq-17}), we can determine the elastic
deformation $\varepsilon _{\rm el}$, at which the lubricant
solidifies,
%20
\begin{equation}
\varepsilon _{{\rm el},c}^{0}=\sqrt{\frac{2\alpha (T_{c}-T)}{a}}.
\label{eq-19}
\end{equation}%
Note that, according to relation~(\ref{eq-18}), the melting can occur
even at the zero temperature, $T=0$, if the strain exceeds the
critical value. At the zero strain, i.e. at $\varepsilon _{\rm
el}=0$, the lubricant melts, when its temperature exceeds the
critical value $T_{c0}$ (see\ Eq.~(\ref{eq-16})).

As a rule, it is the relative shear velocity between friction
surfaces rather than the shear strain component $\varepsilon _{\rm
el}$ that is registered in experiments~\cite{Yosh,Yosh-1996}.
Therefore, for our research to go further, it is necessary to obtain
a relation between those two quantities. Let us take advantage of
the Debye approximation, according to which the elastic strain
component $\varepsilon _{\rm el}$ arises in the lubricant layer,
when the latter flows plastically at the velocity~\cite{Popov}
%21
\begin{equation}
\dot{\varepsilon}_{\rm pl}=\frac{\varepsilon _{\rm el}}{\tau
_{\varepsilon }}, \label{eq-20}
\end{equation}%
where $\tau _{\varepsilon }$ is the Maxwell relaxation time for
internal stresses. The total strain in the layer is determined as
the sum of elastic, $\varepsilon _{\rm el}$, and plastic,
$\varepsilon _{\rm pl}$, components~\mbox{\cite{Popov,Anael2}}
%22
\begin{equation}
\varepsilon =\varepsilon _{\rm el}+\varepsilon _{\rm pl}.
\label{eq-21}
\end{equation}%
Combining relations~(\ref{eq-8}), (\ref{eq-20}), and (\ref{eq-21}),
we obtain the kinetic equation for the evolution of the elastic
component of the shear strain~\cite{1-order,JtfL2,JtfL3,TribInt}:
%23
\begin{equation}
\tau _{\varepsilon }\dot{\varepsilon} _{\rm el}=-\varepsilon _{\rm
el}+\frac{(V-v)\tau _{\varepsilon }}{h}.  \label{eq-22}
\end{equation}

Boundary friction experiments testify that the relaxation time of
the elastic strain is very short, as a rule. This quantity can be
estimated from the relation ${\tau _{\varepsilon }\approx a/c\sim
10^{-12}~}\mathrm{s}$, where ${a\sim 1}$\textrm{~nm} is the lattice
constant or the intermolecular distance, and ${c\sim
10^{3}~}\mathrm{m/s}$ is the sound velocity~\cite{JPS}. However, in
the boundary mode, the strain relaxation time, $\tau
_{\varepsilon }$, can differ by several orders of
magnitude~\cite{Yosh,Yosh-1996}. Bearing in mind that the value of
strain relaxation time $\tau _{\varepsilon }$ is small, below we use
the adiabatic approximation $\tau _{\varepsilon
}\dot{\varepsilon}_{\rm el}\approx 0$~\cite{Olemskoi}, which allows
us to determine the strain by its stationary value
%24
\begin{equation}
\varepsilon _{\rm el}^{0}=\frac{(V-v)\tau _{\varepsilon }}{h},
\label{eq-28}
\end{equation}rather than by Eq.~(\ref{eq-22}).

In the general case, the free energy~(\ref{eq-12}) depends on the
lubricant layer thickness $h$~\cite{2000-SolStCom-Popov}. Note that,
in the framework of our model, the second term in
expression~(\ref{eq-12}) is proportional to the square of the elastic
strain, $\varepsilon _{\rm el}^{2}$. In accordance with
relation~(\ref{eq-28}), the stationary elastic strain increases with
a reduction of the lubricant thickness $h$. Therefore, in the limiting
case of a very thin layer ($h\rightarrow 0$), the strain $\varepsilon
_{\rm el}\rightarrow \infty $. In this case, the second term in
expansion~(\ref{eq-12}) dominates, and the stationary value of order
parameter equals zero, so that the lubricant is liquid-like, as in
work~\cite{2000-SolStCom-Popov}. A detailed study of the influence
of the lubricant layer thickness on friction modes was carried out in
works~\cite{2008-Braun,Aranson}.

\section{Kinetics of Melting}

The changes in the lubricant temperature $T$ and the strain
$\varepsilon_{\rm el} $ induce variations of the order parameter
$\varphi$, which governs the free energy $f$ (see Eq.~(\ref{eq-12}))
in accordance with the power-law expansion of the
latter~\cite{stat}. The stabilization time for a new stationary
value $\varphi_{+}$ (see Eq.~(\ref{eq-15})) is determined by the
generalized thermodynamic force ${-\partial f/\partial\varphi}$. If
${\varphi \approx\varphi_{+}}$, this force is small, and the
relaxation process is described by the Landau--Khalatnikov linear
kinetic equation~\cite{coll}
%25
\begin{equation}
\dot{\varphi}=-\delta\frac{\partial f}{\partial\varphi},   \label{eq-23}
\end{equation}
where the kinetic coefficient $\delta$ characterizes the inertial properties
of the system. After substituting energy~(\ref{eq-12}) into
Eq.~(\ref{eq-23}), we obtain the equation in the explicit form,
%26
\begin{equation}
\dot{\varphi}=-\delta\left(
2\alpha(T-T_{c})\varphi+a\varphi\varepsilon _{\rm
el}^{2}-b\varphi^{2}+c\varphi^{3}\right) +\xi(t).   \label{eq-24}
\end{equation}

Equation (\ref{eq-24}) includes a term responsible for additive fluctuations
with a low intensity~\cite{JtfL2,JtfL3}. Their intensity is selected to be
so low that they do not affect the deterministic behavior of the system.
However, their introduction is necessary, because, at subsequent numerical
calculations, the root $\varphi =0$ of Eq.~(\ref{eq-24}) corresponding to
the maximum of the potential $f(\varphi )$, i.e. to the unstable stationary
state, turns out stable according to the structure of the equation. In this
situation, the introduction of $\xi (t)$ stimulates the system to transit
from the unstable state into a stable one, which corresponds to the energy
minimum. Hence, fluctuations are taken into consideration by means of the
features in subsequent numerical calculations.

The dynamic characteristics of any tribological system are governed
by its properties in whole. For instance, in the geometry
illustrated in Fig.~1, the behavior of the system substantially
depends on the stiffness constant, $K$, of the spring and the mass of the lower
block, $M$. In contrast to the case of motion with constant elastic
strains, this tribological system can reveal the stick-slip mode of
motion in the course of
friction~\cite{Yosh,Yosh-1996,Filippov,period}. The indicated mode
is established because the lubricant periodically melts and
solidifies in the course of motion, which leads to the oscillatory
character of friction force $F$. To calculate the evolution of this
system in time, we need to solve the system of kinetic equations
(\ref{eq-3}) and (\ref{eq-24}) numerically, determining the friction
force $F$ from Eqs.~(\ref{F}) and (\ref{eq-11}), the elastic
stresses $\sigma _{\rm el}$ from Eq.~(\ref{eq-13}), and the strain
$\varepsilon _{\rm el}$ from relation~(\ref{eq-28}). In so doing, we
have to take into account the relation $\dot{x}=v$, as well as
definitions~(1) and~(2).

While solving the differential equations numerically, we used the
Euler--Cramer method with the time increment $\Delta
t=10^{-10}~\mathrm{s.}$ The initial conditions $\varphi
_{0}=x_{0}=v_{0}=0$ were chosen. The result obtained is shown in
Fig.~3. The dashed curve in the upper panel corresponds to the time
dependence of the upper block coordinate $X(t)$ (see Eq.~(1)), and
the solid one to that of the lower block, $x(t)$, which is more
complicated. The figure also exhibits the time dependences for the
block velocities, elastic shear stresses $\sigma _{\rm el}$ (see
Eq.~(\ref{eq-13})) that arise in the lubricant, and total friction
force $F$ (see Eq.~(\ref{F})). Let us examine these dependences in
more details.

%Fig. 3
\begin{figure}% figure* for wide figure, [h] [!] to change the placement
\includegraphics[width=7cm]{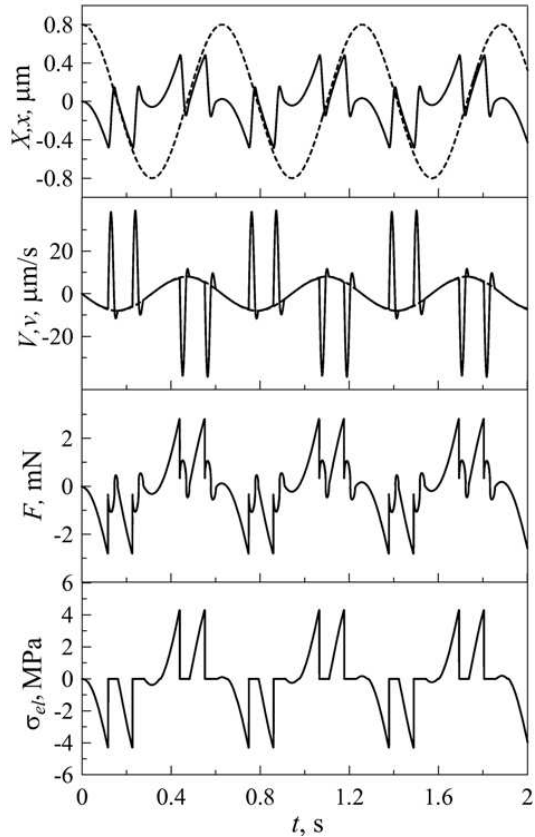}
\vskip-3mm\caption{Dependences of the coordinates $X$ and $x$, the
velocities $V$ and $v $, the elastic stresses $\sigma_{\rm el}$ (see
Eq.~(\ref{eq-13})), and the friction force $F$ (see Eq.~(\ref{F}))
on the time $t$ for the same parameters as in Fig.~2 and
${h=10^{-9}}$~m, ${\tau_{\varepsilon}=10^{-8}}$~s, ${\gamma=-2/3}$,
${A=0.6\times10^{-9}}$~m$^{2}$,
${k=5\times10^{4}}$~$\mathrm{Pa}\cdot\mathrm{s}^{1/3}$,
${\delta=100}$~m$^{3}/(\mathrm{J\cdot s})$, ${T=200}$~K,
${X_{m}=0.8\times10^{-6}}$~m, ${\omega=10}$~rad/s, ${M=0.4}$~kg, and
${K=3000}$~N/m. The dashed curves correspond to the coordinate
$X(t)$ and the velocity $V(t)$ of the upper block, and the solid
ones to the coordinate $x(t)$ and the velocity $v(t)$ of the lower
block  }
\end{figure}

At the initial time moment, $t=0$, the blocks are motionless, and
the lubricant is in the solid-like state, because the dependences
are plotted for the lubricant temperature $T$ lower than the
critical one, $T_{c}^{0}$ (see Eq.~(\ref{eq-17})), and $\varepsilon
_{\rm el}=0$ at rest. At $t=0$, the upper block starts to move, and,
at $t>0$, its velocity grows in accordance with Eq.~(2). Since the
lubricant is in the solid-like state, the friction force $F$
possesses both the viscous and elastic components, and the lower
block moves together with the upper one. However, in the course of
motion, the absolute value of elastic force $2Kx$, which impedes the
lower block to move, grows and, as a result, the velocity $v$ does
not increase so sharply as the velocity $V$ does. Hence, the
relative shear velocity between block surfaces, $V-v$, increases in
time and, in accordance with Eq.~(\ref{eq-28}), the elastic strain
$\varepsilon _{\rm el}$ also grows. At a certain time moment, the
condition $\varepsilon _{\rm el}>\varepsilon _{{\rm el},c0}$ (see
Eq.~(\ref{eq-18})) becomes satisfied, and the lubricant begins to
melt following the \textquotedblleft shear
melting\textquotedblright\ mechanism. The friction force
substantially decreases at that, because the stresses vanish, and
the lower block can slide for a considerable distance, being driven
by the elastic force from the compressed and stretched springs.
Therefore, the relative shear velocity diminishes, and, when the
condition $\varepsilon _{\rm el}<\varepsilon _{{\rm el},c}^{0}$ (see
Eq.~(\ref{eq-19})) is obeyed, the lubricant solidifies again. The
considered process repeats periodically.

%Fig. 4
\begin{figure}% figure* for wide figure, [h] [!] to change the placement
\includegraphics[width=7cm]{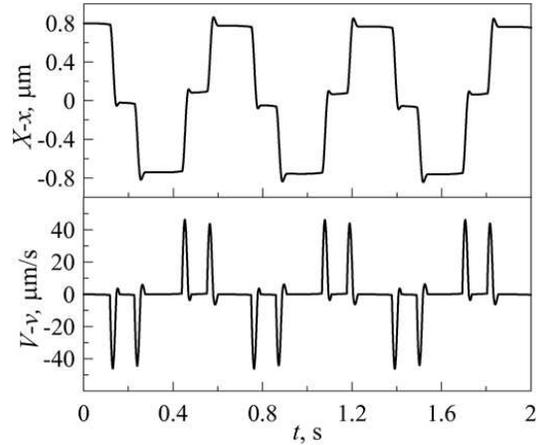}
\vskip-3mm\caption{Dependences of the relative displacement, $X-x$,
and the velocity, $V-v$, of blocks on the time $t$ for the same
parameters as in Fig.~3  }
\end{figure}

In addition, Fig.~4 demonstrates the time dependences of the relative block
displacement and the velocity. At the time moments, when the surfaces
\textquotedblleft stick\textquotedblright\ to each other, their relative
displacement $X-x$ remains constant, and the relative shear velocity $V-v$
is close to zero (in this case, the dependences $V(t)$ and $v(t)$ in Fig.~3
visually coincide). Hence, the periodic stick-slip mode of motion takes
place, which is also typical of dry friction, when no lubricant is
used~\cite{Persson,PopovBook,trib-system}. For the chosen parameter values, the blocks
\textquotedblleft stick\textquotedblright\ to each other four times during a
complete period of parameter changes: two times in each direction of motion,
with the obtained dependences being symmetric with respect to the motion
direction. However, a number of different situations can be realized
depending on the system parameters.

%Fig. 5
\begin{figure*}% figure* for wide figure, [h] [!] to change the placement
\includegraphics[width=11.6cm]{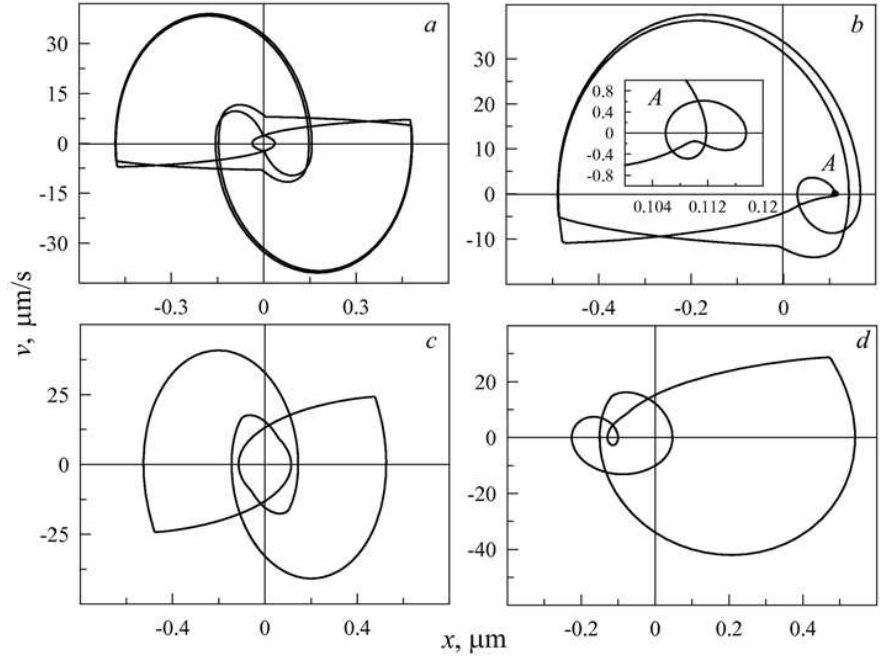}
\vskip-3mm\caption{Phase portraits of the system at the same
parameters as in Fig.~3 and various values of the cyclic frequency
$\omega=10$ ({\it a}), 15 ({\it b}), 32 ({\it c}), and 38~rad/s
({\it d}) }
\end{figure*}

The phase portraits of the system calculated at the same parameters as in
Fig.~3 and various values of cyclic frequency $\omega $ are depicted in
Fig.~5. The kinetic dependences in Fig.~3 completely correspond to the phase
portrait in Fig.~5,$a$, because they were calculated for the same frequency
$\omega $ value. It is important to emphasize the fact that the phase
portraits in Figs.~5,$a$ and $c$ are symmetric with respect to the
coordinate origin, whereas the phase portraits in Fig.~5,$b$ and $d$
illustrate the situation where the motion of the upper friction surface in
one direction affects differently the motion of the lower block in
comparison with its motion in the opposite direction. Hence, the system
reveals memory effects, which were observed experimentally~\cite{Yosh}. In
this case, the motion of the lower block is also periodic in time, but the
time dependences of the parameters, which are given in Fig.~3, are not symmetric
with respect to their zero values~\cite{JtfL2}. The inset in Fig.~5,$b$
demonstrates the enlarged section conditionally marked by letter A, because
this section has pronounced features, which cannot be distinguished in the
main plot. Thus, the frequency $\omega $ affects the behavior of
the tribological system in a non-trivial manner. By varying $\omega $, it is
possible to select various modes of motion, which considerably differ from
one another. Note that, at some frequencies, the stationary behavior of the
system, which is established as a result of the system evolution, depends on the
initial conditions or the system prehistory. For instance, in Fig.~5,$d$,
the initial value $\varphi _{0}\neq 0$ gives rise to a mode similar to that
exhibited in Fig.~5,$c$. This circumstance also confirms the presence of
memory effects in the system, which were observed experimentally~\cite{Yosh}.

%Fig. 6
\begin{figure}% figure* for wide figure, [h] [!] to change the placement
\includegraphics[width=7.05cm]{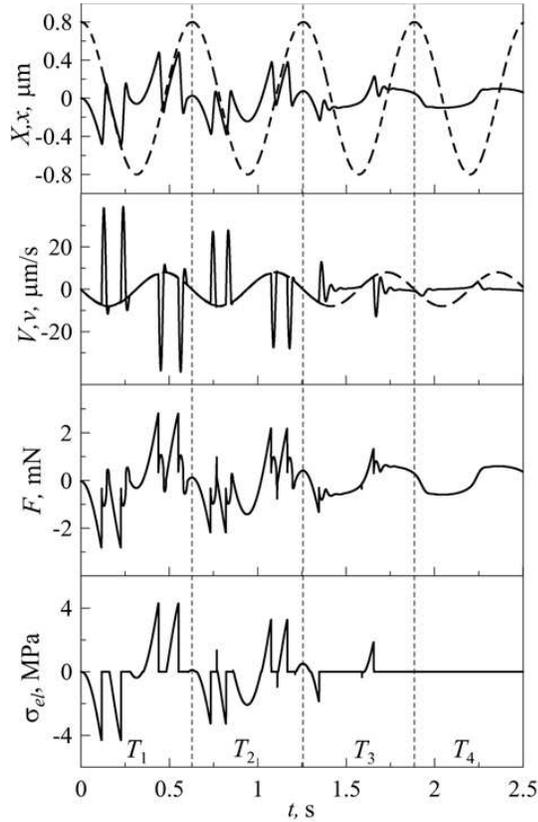}
\vskip-3mm\caption{Dependences of the coordinates $X$ and $x$, the
velocities $V$ and $v $, the elastic stresses $\sigma_{\rm el}$ (see
Eq.~(\ref{eq-13})), and the friction force $F$ (see Eq.~(\ref{F}))
on the time $t$ for the same parameters as in Fig.~3 and the
temperatures $T_{1}=200$~K, $T_{2}=220$~K, $T_{3}=250$~K, and
$T_{4}=300$~K${.} $The dashed curves correspond to the $X(t) $ and
$V(t)$ dependences, and the solid curves to the $x(t)$ and $v(t)$
ones  }\vskip-2mm
\end{figure}

%Fig. 7
\begin{figure}% figure* for wide figure, [h] [!] to change the placement
\includegraphics[width=8cm]{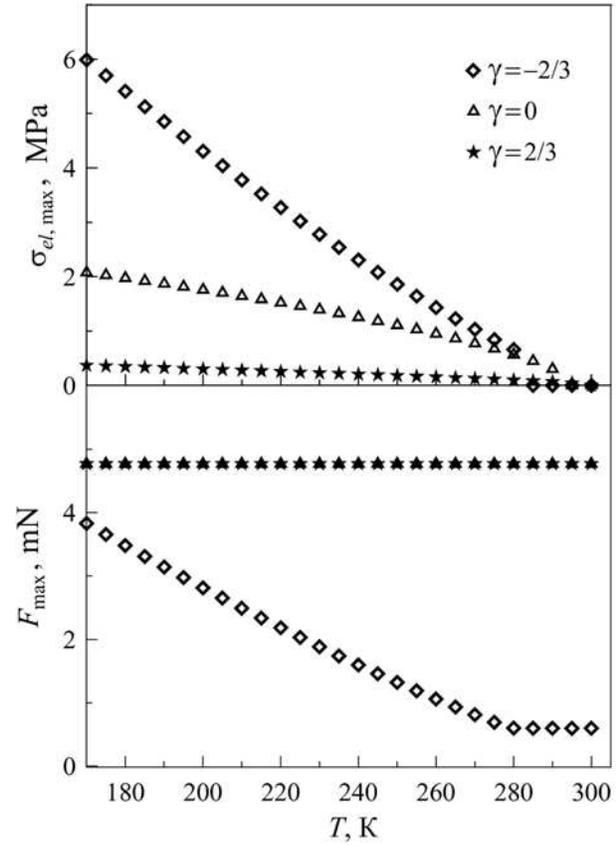}
\vskip-3mm\caption{Dependences of the elastic stress, $\sigma_{{\rm
el},\mathrm{max}}$, and friction force, $F_{\mathrm{max}}$,
amplitudes on the temperature $T$ for pseudoplastic
($\gamma=-2/3$), Newtonian ($\gamma=0$), and dilatant ($\gamma=2/3$)
fluids as a lubricant. The parameters are the same as in Fig.~3  }
\end{figure}

Figure~6 elucidates the influence of lubricant temperature $T$ on
the melting kinetics. The plotted dependences are divided into four
sections. The temperature for the first section is the lowest, and,
for every next section, the temperature increases, i.e. we have the
inequalities $T_{1}<T_{2}<T_{3}<T_{4}$. The dependence obtained in
the first section, at $T=T_{1}$, reproduces the dependence shown in
Fig.~3 in more details, because it was obtained at the same
$T$-value. As the temperature is elevated to $T=T_{2}$, the
stick-slip mode of motion is realized, as it was at $T=T_{1}$.
However, the maximum value of elastic stresses $\sigma _{\rm el}$
decreases at $T=T_{2}$. As a result, the friction force $F$ in the
solid-like lubricant also decreases, as the temperature grows. As
the temperature is elevated to $T=T_{3}$, this tendency survives.
Note that a reduction of the sticking peak number with the
temperature growth is not a rule, and the opposite situation can
take place. At $T=T_{4}$, the lubricant is liquid-like all the time,
and the elastic stresses equal zero. It is so, because, at this
temperature, the condition $T>T_{c}^{0}$ (see Eq.~(\ref{eq-17})) is
obeyed even if $\varepsilon _{\rm el}=0$, i.e. the melted lubricant
cannot solidify due to a reduction of the relative shear velocity
between the rubbing surfaces. We do not know of any experiments
devoted to similar researches of the influence of the temperature on
the friction mode. Therefore, the dependences exhibited in Fig.~6
are a forecast.\vspace*{-1.5mm}

\section{Numerical Experiment}

The dependences shown in Fig.~6 testify that the growth of the
temperature $T$ gives rise to a reduction of the elastic stress
amplitude $\sigma _{\rm el}$ and a reduction of the friction force
$F$ maximum. Let is analyze the dependences of the $\sigma _{\rm
el}$ and $F$ amplitudes on the temperature $T$ at various modes of
functioning of the system in more details. We define the stress
amplitude as ${\sigma _{{\rm el},\mathrm{max}}:=\left( \sigma _{{\rm
el},\mathrm{max}}-\sigma _{{\rm el},\mathrm{min}}\right) /2}$ and
the friction force amplitude as ${F_{\mathrm{max}}:=\left(
F_{\mathrm{max}}-F_{\mathrm{min}}\right) /2}$, where $\sigma _{{\rm
el},\mathrm{max}}$ and $F_{\mathrm{max}}$ are the maximum values of
elastic stresses and friction force, respectively, and $\sigma
_{{\rm el},\mathrm{min}}$ and $F_{\mathrm{min}}$ are their minimum
values, which are determined within the complete period of parameter
changes, $T=2\pi /\omega $, after the stationary friction mode has
been established.

%Fig. 8
\begin{figure}% figure* for wide figure, [h] [!] to change the placement
\includegraphics[width=7.0cm]{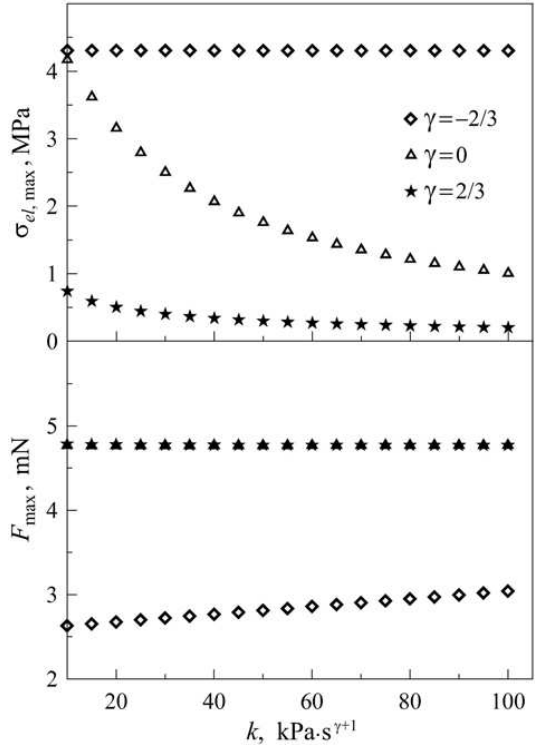}
\vskip-3mm\caption{Dependences of the elastic stress, $\sigma_{{\rm
el},\mathrm{max}}$, and friction force, $F_{\mathrm{max}}$,
amplitudes on the proportionality coefficient $k$ (see
Eq.~(\ref{eq-7})) for pseudoplastic ($\gamma=-2/3$), Newtonian
($\gamma=0$), and dilatant ($\gamma=2/3$) fluids as a lubricant. The
parameters are the same as in Fig.~3  }\vskip-1.5mm
\end{figure}

The dependences of the indicated quantities on the temperature are
depicted in Fig.~7 for three types of lubricants: pseudoplastic
($\gamma <0)$, Newtonian ($\gamma =0$), and dilatant ($\gamma >0$)
fluids. The upper panel demonstrates that, as the temperature
increases, the elastic stresses $\sigma _{{\rm el},\mathrm{max}}$
decrease for all three types of fluids, i.e. the temperature
elevation favors the lubricant melting. Note that, for pseudoplastic
fluids ($\gamma =-2/3$), which are used most often as lubricants in
such systems, the stress amplitude attains maximum values within
almost the whole presented range of temperatures, but the melting occurs
at lower $T$ in this case. The lower panel of the figure shows the
dependences of the friction force amplitudes $F_{\mathrm{max}}$ on the
lubricant temperature $T$. It follows from the figure that the
friction force decreases with the temperature growth only for
pseudoplastic fluids, and it is minimal within the whole range of
temperatures in comparison with other types of fluids. For dilatant
and Newtonian fluids and for the selected parameter values, the
maximum friction force does not change with the temperature growth.
Since the elastic stresses for those fluids decrease as the temperature
grows (the upper panel of the figure), this means that the growth of
$T$ gives rise to an increase of the viscous component of the friction
force, to which the second term in formula~(\ref{F}) corresponds.
In the situation concerned, this can happen only if the relative
velocity of motion, $V-v$, increases. Note that, according to
the figure, the amplitude of the friction force for the Newtonian and
dilatant fluids remains constant when the temperature grows, even in
the case $\sigma _{\rm el}=0$, i.e. when the friction force has only
the viscous component. Since the $F$-amplitudes in the cases $\gamma
=0$ and $\gamma =2/3$ coincide at all temperatures, it is not
sufficient to experimentally measure the total friction force in
order to determine the friction mode. That is why the behavior of
the elastic, $\sigma _{\rm el}$, and viscous, $\sigma _{v}$, stresses
are additionally studied as a rule~\cite{exp-n}. Note also that,
according to the results demonstrated in Fig.~7, the application of
pseudoplastic fluids is optimal to reduce friction, because they
favor the establishment of a mode with minimum force $F$, despite
that the elastic stresses for such lubricants are maximum within almost
the whole range of temperatures.

%Fig. 9
\begin{figure}% figure* for wide figure, [h] [!] to change the placement
\includegraphics[width=7.1cm]{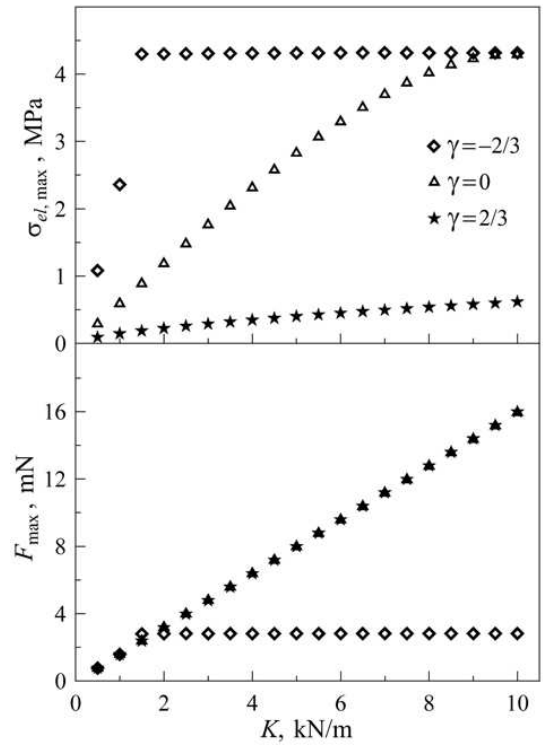}
\vskip-3mm\caption{Dependences of the elastic stress, $\sigma_{{\rm
el},\mathrm{max}}$, and friction force, $F_{\mathrm{max}}$,
amplitudes on the spring stiffness constant $K$ (see
Eq.~(\ref{eq-3})) for pseudoplastic ($\gamma=-2/3$), Newtonian
($\gamma=0$), and dilatant ($\gamma=2/3$) fluids as a lubricant. The
parameters are the same as in Fig.~3  }\vspace*{-1mm}
\end{figure}

To determine the dependence of the viscosity on the velocity
gradient and the temperature, both real~\cite{Wear} and
computer-assisted~\cite{Perss-PRL} experiments are carried out. The
problem urgency is connected with the fact that the dependences of
the viscosity on the indicated quantities are anomalous in the
boundary friction mode in the case of nano-sized tribological
systems. There can even be a mode, when the friction force almost
vanishes at cryogenic temperatures, which corresponds to a low
viscosity of a lubricant and, accordingly, a very weak energy
dissipation. In the English-language scientific literature, this
mode was coined as \textquotedblleft
superlubricity\textquotedblright~\cite{super1,super2}. Let us
examine the dependences of the friction force and the stresses for
three types of lubricants -- however, not on the temperature (as in
Fig.~7), but on the proportionality coefficient $k$ between the
viscosity and the velocity gradient (see Eq.~(\ref{eq-7})). The
corresponding plots are depicted in Fig.~8. Note that, in contrast
to Fig.~7, different $k$-values correspond to different lubricants,
friction surfaces, or experimental geometries. This means that every
point in the dependences exhibited in Fig.~8 corresponds to
tribological systems different by their properties. As one can see,
for pseudoplastic fluids ($\gamma=-2/3$), the elastic stresses
remain constant with increase of the coefficient $k$. For Newtonian
and dilatant fluids, the maximum stresses monotonously decrease with
the increase of $k$. The friction force amplitude $F_{\mathrm{max}}$
grows with the coefficient $k$ in the case of pseudoplastic fluid
($\gamma=-2/3$). At the same time, for the indices $\gamma=0$ and
$\gamma=2/3$, the friction force behaves identically as in Fig.~7,
i.e. it remains constant. However, within the whole presented range
of $k$-values, $F_{\max}$ is minimal just for the pseudoplastic
fluid; therefore, the latter is optimal for creating the conditions
to reduce the friction in this case~as~well.\looseness=1

In Fig.~9, the behavior of the examined quantities is illustrated, as the spring
stiffness constant $K$ increases. For the dilatant and
Newtonian fluids, the elastic stresses $\sigma _{{{\rm
el},\mathrm{max}}}$ monotonously and slowly grow. In the case of
the pseudoplastic fluid ($\gamma =-2/3$), the stresses drastically
increase firstly, and afterward remain almost constant. The friction
force in this case ($\gamma =-2/3$) also grows to a certain value
and, then, does not almost change. For the indices $\gamma =0$ and
$\gamma =2/3$, the amplitudes of friction force $F_{\mathrm{max}}$
linearly increase with the spring stiffness constant $K$,
and their magnitudes are equal as in the previous two figures.
Hence, in this case, the pseudoplastic fluid also provides the
minimum friction force in the system. Thus, a general conclusion can
be drawn that the pseudoplastic fluids provide an optimal friction mode
in the tribological system exhibited in Fig.~1, because the maximum
friction force $F_{\mathrm{max}}$ is the lowest for them.

\section{Conclusions}

In this work, a thermodynamic model was developed to describe the
behavior of a tribological system functioning in the boundary
friction mode. The model allowed a number of effects observed
experimentally to be explained. It was shown that the stick-slip
mode of motion is a result of the phase transition of the first
order between the liquid- and solid-like states of a lubricant. The
influence of the lubricant temperature, the spring stiffness
constant, and the coefficient of proportionality between the
viscosity and the velocity gradient on the system behavior was
analyzed. For pseudoplastic fluids, the elastic stresses and the
friction force were found to decrease with the temperature growth.
The increase of the spring stiffness constant induces the growth of
the friction force and stresses for all types of lubricants. When
the coefficient of proportionality $k$ increases, the maximum
stresses do not change substantially in the case of pseudoplastic
fluids, whereas the friction force grows. For the sake of
comparison, the results of calculations obtained for the dilatant
and Newtonian fluids were also reported. Modes, in which the
displacement between the friction surfaces does not correspond to
the direction of motion of the upper block, were revealed, which
evidences the presence of memory effects in the system. While
developing the model, the thermodynamic potential with two stable
stationary states was used, in which the zero and nonzero minima
were separated by a maximum. However, it was found experimentally
that the lubricant is characterized by more than one type of
transition and it can exist in a few (solid- or liquid-like)
metastable states. For such a situation to be described, the
additional terms of higher orders in the free energy expansion are
sufficient to be taken into account.

\vskip3mm

I.A. Lyashenko is grateful to Prof.~B.N.J.~Persson for his
invitation to make a research visit to the Forschungszentrum
(J$\mathrm{\ddot{u}}$lich, Germany), with this work being partially
fulfilled there. He also thanks the organizers of the Joint
ICTP-FANAS Conference on Trends in Nanotribology (September 12--16,
2011, Miramare, Trieste, Italy) for their invitation and financial
support, as well as to A.E.~Filippov and V.N.~Samoilov for the
discussion of this work at the indicated conference.

The work was supported by the the Ministry of Education and Science, Youth
and Sport of Ukraine in the framework of the project \textquotedblleft
Modeling of friction for metal nanoparticles and boundary liquid films
interacting with atomically smooth surfaces\textquotedblright\
(N~0112U001380).

\vskip-3mm
\rezume{Я.О.~Ляшенко, А.М.~Заскока}{%
ПЕРЕРИВЧАСТИЙ РЕЖИМ МЕЖОВОГО ТЕРТЯ \\ЯК ФАЗОВИЙ ПЕРЕХІД ПЕРШОГО
РОДУ} {Розглянуто трибологічну систему, що складається з двох
контактуючих блоків, один з яких закріплений між двома пружинами, а
інший приведений в неперервний періодичний рух. Досліджено кінетику
системи в режимі межового тертя, коли між атомарно-гладкими
поверхнями блоків знаходиться ультратонка плівка мастила. Для опису
стану мастила записано вираз для густини вільної енергії у вигляді
розкладання в ряд за степенями параметра порядку, який зводиться до
модуля зсуву. Показано, що в широкому діапазоні параметрів
реалізується переривчастий режим руху, до якого приводять періодичні
фазові переходи першого роду між кінетичними режимами тертя.
Спрогнозовано поведінку системи при зміні зовнішніх та внутрішніх
параметрів.}

\end{document}